\documentclass[aps, reprint, prl, twocolumn, superscriptaddress,amsmath,amssymb, longbibliography,noeprint]{revtex4-2}

\usepackage{times}
\usepackage{amsmath}
\usepackage{amssymb}
\usepackage{xfrac}
\usepackage{dsfont}
\usepackage{graphicx} 
\usepackage{float}
\usepackage{dcolumn} 
\usepackage{bm} 
\usepackage{braket}
\usepackage{mathtools}
\usepackage{xcolor}
\setcitestyle{round}
\usepackage[resetlabels,labeled]{multibib}
\newcites{SM}{References}

\newcommand{\affphys}{Department of Physics and Astronomy, Seoul National University, Seoul 08826, Korea}
\newcommand{\affcces}{Center for Correlated Electron Systems, Institute for Basic Science, Seoul 08826, Korea}
\newcommand{\affiap}{Institute of Applied Physics, Seoul National University, Seoul 08826, Korea}
\newcommand{\tqq}{t_\textrm{q}}
\newcommand{\nd}{N_\textrm{d}}
\newcommand{\nv}{N_\textrm{v}}
\newcommand{\akz}{\alpha_\textrm{KZ}}
\newcommand{\kf}{k_\textrm{F}}
\newcommand{\uc}{U_\textrm{c}}
\newcommand{\uf}{U_\textrm{f}}

\begin{document}

\title{Observation of universal Kibble--Zurek scaling in an atomic Fermi superfluid}

\author{Kyuhwan Lee}
\affiliation{\affphys}
\affiliation{\affcces}
\author{Sol Kim}
\affiliation{\affphys}
\affiliation{\affcces}
\author{Taehoon Kim}
\affiliation{\affphys}
\author{Y. Shin}
\affiliation{\affphys}
\affiliation{\affcces}
\affiliation{\affiap}


\begin{abstract}
Half a century ago, T. Kibble proposed a scenario for topological defect formation from symmetry breaking during the expansion of the early Universe.
W. Zurek later crystallized the concept to superfluid helium, predicting a power-law relation between the number of quantum vortices and the rate at which the system passes through the lambda transition.
Here, we report the observation of Kibble--Zurek scaling in a homogeneous, strongly interacting Fermi gas undergoing a superfluid phase transition.
We investigate the superfluid transition using two distinct control parameters: temperature and interaction strength. 
The microscopic physics of condensate formation is markedly different for the two quench parameters, signaled by their two orders of magnitude difference in the condensate formation timescale.  
However, regardless of the thermodynamic direction in which the system passes through a phase transition, the Kibble--Zurek exponent is identically observed to be about 0.68 and shows good agreement with theoretical predictions that describe superfluid phase transitions. 
This work demonstrates the gedanken experiment Zurek proposed for liquid helium that shares the same universality class with strongly interacting Fermi gases.   
\end{abstract}

\maketitle


A fundamental question in physics is how a system can transition into an ordered phase with broken symmetry. The Kibble--Zurek (KZ) mechanism describes the spontaneous formation of topological defects that occur in non-equilibrium phase transitions \cite{Kibble, Zurek}. 
When a control parameter is changed across a continuous phase transition, near the critical point, the system's evolution is effectively frozen due to the critical slowing down.
As a result, the frozen patches of correlated spatial fluctuations pass on to the development of the order parameter.
After the causally independent phase domains appear and merge, the unresolved singularities at domain boundaries remain as topological defects. 

Scale invariance lies at the heart on why apparently different systems display universality near a continuous phase transition.
The length- ($\xi$) and timescale ($\tau$) that govern the universal singular properties near a critical point diverge as $\xi\sim\left\vert\lambda\right\vert^{-\nu}$ and $\tau\sim\left\vert\lambda\right\vert^{-\nu z}$, where $\lambda$ is the distance from the continuous phase transition.
The critical exponents $\nu$ and $z$ are entirely given by the macroscopic system parameters such as the symmetry, dimension, range of interaction, and the presence of conservation laws or additional slow variables \cite{Kardar, H&H}. 
A central prediction of KZ mechanism that originates from scale invariance is the power-law relation, 
\begin{equation}
\nd \propto \tqq^{-\akz},
\end{equation}
between the average number of defects ($\nd$) and quench time ($\tqq$).
The KZ exponent ($\akz$) is universally determined by the critical exponents, $\nu$ and $z$, as well as by the dimensions of the system and defects.

Four decades ago, as a representative example of a spontaneously broken continuous symmetry, Zurek originally considered a normal-to-superfluid transition \cite{Zurek, Zurek2} and predicted the KZ exponent to be
\begin{equation}
\akz = {2\nu}/{(1+\nu z)}.
\end{equation}
Superfluid phase transitions, with its spontaneously broken global $U(1)$ symmetry, have profound implications in the study of quantum liquids \cite{Leggett}.
The density of relevant quantum liquids span magnitudes of order, with the list including liquid heliums ($^\textrm{3}$He, $^\textrm{4}$He), superconductors, ultracold atomic gases and a possible extension to neutron stars.
Measurements of the critical exponents $\nu$ and $z$ have shed light on the scale-invariant properties near a superfluid transition \cite{Goldner, Donner, Navon}.
However, verifying the scaling behavior of defect formation dynamics in a superfluid phase transition (Eqs. 1 and 2) remains elusive \cite{Hendry, Bauerle, Ruutu, Dodd, Weiler, Lamporesi, Donadello, Chomaz, Ko, Liu, Rabga}.
Major challenges include fast relaxation of turbulence (or the slowness of mechanical quenches) in $^\textrm{4}$He \cite{Dodd, Karra}, limited dynamic range of a control parameter in $^\textrm{3}$He \cite{Bauerle}, density inhomogeneity \cite{Donadello, Ko, Liu, Rabga} and finite-size effects \cite{Chomaz} in ultracold atomic gases.
For a recent review on the topic, see Refs. \cite{delCampo1, Beugnon}.

\begin{figure*}
\includegraphics{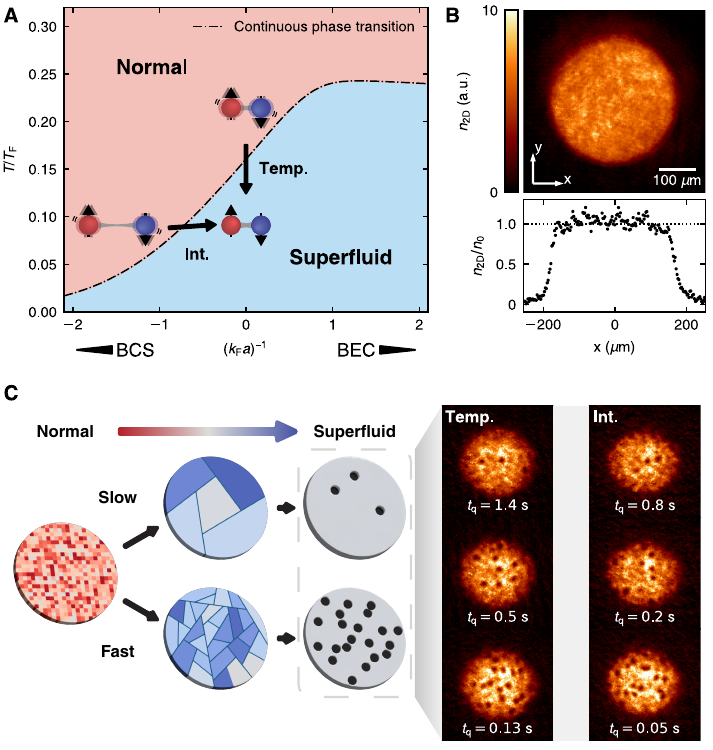}
\caption
{
\textbf{Spontaneous defect formation in a homogeneous atomic Fermi superfluid.}
(\textbf{A}) Phase diagram of a strongly interacting Fermi gas in the Bardeen--Cooper--Schrieffer (BCS)--Bose--Einstein condensate (BEC) crossover~\cite{Haussmann} and the normal-to-superfluid quench trajectories studied in this work.
The vertical (horizontal) arrow indicates the temperature (interaction) quench.
In the interaction quench, the temperature increases slightly due to the isentropic tuning of the interaction parameter \cite{Carr, Dyke}.
(\textbf{B}) Column density image of a unitary Fermi gas in the superfluid phase, averaged over 5 different experimental realizations. 
A horizontal line cut along the center of the sample is shown below.
$n_\textrm{0}$ is the average column density within the center disk region of a diameter of $310~\mu\text{m}$.
(\textbf{C}) Schematic of the experiment.
After quenching a control parameter (temperature or interaction strength) over a time $\tqq$ through the superfluid phase transition, the quantum vortices created in the sample are detected with their density-depleted cores in time-of-flight (ToF) imaging.
}
\end{figure*}

Here, we report the observation of KZ scaling in a homogeneous, strongly interacting Fermi gas undergoing a superfluid phase transition.
We employ two different quench parameters: temperature and interaction strength (Fig. 1A).  
Compared to earlier experiments in ultracold atomic systems that relied on macroscopic temperature quench, the magnetic Feshbach resonance between spin up and spin down fermions offers a new opportunity to examine the critical dynamics by tuning the microscopic interaction strength.  
Regardless of the different timescales involved in condensate formation, for both quench parameters, we observe an identical power-law behavior that is indicative of the universal nature of continuous phase transitions.
Furthermore, we find that the power-law exponent for the slow quench regime is accurately described within the two-fluid hydrodynamics of model $F$, in which the predicted value is 0.67.


Our experiment starts by preparing a thermal cloud of $^{6}$Li atoms in a balanced mixture of the two lowest hyperfine states.
The Feshbach magnetic field is tuned to 830 G so that the atomic cloud is in the unitary regime with $(\kf a)^{-1} = 0$, where $\kf$ is the Fermi wavenumber and $a$ is the $s$-wave scattering length.
Initially, the atoms are trapped in an oblate trap potential, where the optical dipole trap (ODT) gives the tight confinement along the vertical direction and the coils that generate the Feshbach magnetic field give the dominant radial harmonic potential.
The tight confinement along the vertical direction ensures that independent phase domains are created only along the plane perpendicular to the line of sight. 
Before initiating a quench, we slowly ramp up a repulsive 532 nm laser beam that has been spatially tailored by a spatial light modulator to create an in-plane homogeneous trap potential of a disk geometry (Fig. 1B) \cite{supp}. The sample diameter is about 360~$\mu$m, allowing for a wide dynamic range of defect numbers.

The trap homogenization is essential to meet a fundamental requirement of KZ scaling, that is, the absence of causal connection between the creation of phase domains. 
Systems with trap inhomogeneity experience phase transitions at different times in different locations. This means that the phase information of broken symmetry at one place can spread and influence the phase transitions at other places, thus impeding the formation of defects.
It has been shown that the effect of causality on suppressing defects results in a power-law exponent that is much higher than that of Eq. 2 \cite{delCampo2, Ko, Goo1, Rabga}. However, its quantitative understanding is limited because of a lack of knowledge of the dynamics of inhomogeneous phase transitions.


\begin{figure}
\includegraphics{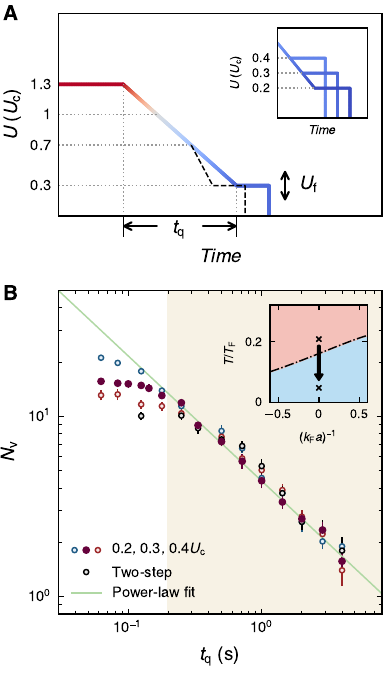}
\caption
{
\textbf{Temperature quench.}
(\textbf{A}) Illustration of various temperature quench protocols. 
The solid line in the main figure shows a single linear quench, where $U$ denotes the depth of the ODT and $\uc$ is the critical depth for the phase transition. 
The inset displays linear quench protocols with different final trap depths ($\uf$).
The black dashed line in the main figure indicates a two-step quench.
(\textbf{B}) Average number of vortices ($\nv$) versus temperature quench time ($\tqq$).
The inset depicts the quench trajectory along the phase diagram. 
The solid data points represent a single linear quench with $\uf= 0.3\uc$.
The blue (red) open circles indicate a single linear quench with $\uf= 0.2\uc$ (0.4$\uc$).
The black open circles show the result for a two-step quench.
The solid line is a power-law fit of the solid data points in the shaded region.
For the fast quench regime (unshaded region), the number of vortices shows a large dependency on the specific quench protocol, suggesting the impact of early-time coarsening during the initial growth of the condensate.
Each data point is an average of 20 to 60 experimental realizations. 
The error bars denote the standard error of the mean.
When the error bars are not visible, they are smaller than the marker size.
}
\end{figure}

We perform a temperature quench by linearly lowering the depth of the ODT. As previously shown in \cite{Ko}, the linear relation between trap depth and momentum distribution widths allows us to use ODT depth as a proxy for temperature. 
The initial number of atoms per spin state at a trap depth of 1.3$\uc$ is 1.0$\times$$10^{6}$, where $\uc$ denotes the critical trap depth at which the unitary Fermi gas undergoes a phase transition.
The vertical trap frequency at $\uc$ is $\omega_{z}=2\pi\times$690 Hz. 
We decrease the trap depth from 1.3$U_{c}$ to its final value of 0.3$\uc$ over a time $\tqq$ ranging from 50~ms to 4~s (Fig. 2A). 
We then wait an additional 0.2~s so that the condensate fraction has reached an equilibrium level larger than $70\%$ \cite{supp}.
The hold time is much shorter than the typical vortex decay time ($\sim 1~\textrm{s}$).
Finally, we release the trap and take a time-of-flight (ToF) image of the atomic cloud to detect any created vortices (Fig. 1C). 
During the ToF expansion, the vortex core size increases, while the vortex line remains parallel to the line of sight.

The solid circles in Fig. 2B show the average number of vortices ($\nv$) as a function of quench time ($\tqq$). 
Spanning over an order of magnitude in quench time, the shaded region in Fig. 2B specifies the scaling regime in which the KZ mechanism accurately describes the critical dynamics of spontaneous defect formation.
For fast quenches with $\tqq<0.2$~s, however, we observe a saturation of the average number of vortices, which we attribute to the early-time coarsening of the order parameter \cite{Chesler,Goo1,Goo2}.
Focusing on the scaling regime, a power-law fit to the number of vortices as a function of quench time  gives $\akz=0.69(2)$ \cite{supp}. According to the so-called model $F$ in a dynamic renormalization group theory, the static and dynamic critical behavior of the order parameter features $\nu\approx0.67$ and $z=1.5$ \cite{Siggia1, Siggia2, H&H}, predicting the KZ exponent (Eq. 2) to be 0.67, which shows good agreement with the measured $\akz$.
This is in stark contrast to the large scaling exponent ($\geq2$) observed in a harmonic trap \cite{Ko}.

To further confirm that KZ scaling is independent of the quench protocol,
we employ different quench protocols after passing through the critical point.
First, the final trap depth ($\uf$) is varied for a linear quench, as shown by the slightly different blue shades in the inset of Fig. 2A.
Also, a two-step quench protocol is adopted, in which the system is linearly quenched in two steps.
At varying rates, the first quench stops at a trap depth of 0.7$\uc$. 
For the remaining second step, the trap depth is lowered to 0.3$\uc$ with a fixed time of 133 ms (dashed line in Fig. 2A), which corresponds to the second fastest quench rate in the scaling regime of Fig. 2B.
Based on recent experimental results of second sound diffusion in homogeneous unitary Fermi gases, we expect that a trap depth of 0.7$\uc$ is shallow enough to be out of the critical freeze-out window \cite{Li, Yan}.

The results of the number of vortices as a function of quench time for various temperature quench protocols are shown in Fig. 2B.
For a direct comparison with the solid-circle data points, the horizontal axis has been adjusted to appropriately represent the quench rates when passing through the critical point.
The blue (red) open circles correspond to $\uf=0.2\uc$ (0.4$\uc$).
The black open circles indicate the results for the two-step quench.
As highlighted in the shaded region of Fig. 2B, the scaling behavior for all the different quench trajectories collapse into a single curve.
This signifies that the power-law behavior in the slow quench regime is independent of the quench protocol and further reinforces our measurement of the KZ exponent. 

The saturation of the defect number for rapid quenching reveals the predominance of coarsening dynamics in the defect formation after the critical freeze-out period.
During the initial nonadiabatic growth of the order parameter, when the presence of a condensate has yet to be established, the incipient spatial fluctuations may be coarsened, thus decreasing the probability of forming quantum vortices \cite{Chesler}.
This idea goes beyond the original concept of KZ theory, which assumes that the long-wavelength modes of the condensate follow an adiabatic growth curve immediately after exiting the freeze-out window. 
The effects of such early-time coarsening are more pronounced for rapid quenches, as evidenced by the observation that the condensate fraction at the end of the quench is much lower than its equilibrium value (Fig. 4) \cite{Goo1,supp}.

The saturation level of the vortex number varies significantly depending on the quench protocol used.
For our fastest quench, the vortex number at $\uf=0.4\uc$ is up to 40$\%$ lower than at $\uf=0.2\uc$.
This reflects that higher temperature translates to a prolonged period of coarsening dynamics.
The lower saturated number of vortices for a two-step quench compared to a single quench further demonstrates the impact of early-time coarsening.
The quench rate through the critical point and the final trap depths are identical for both quench protocols, but a slower second quench time results in a 30$\%$ smaller number of vortices for the two-step quench.
It should be noted that vortex relaxation shows negligible temperature dependence at low temperature \cite{supp}. 
This suggests that vortex relaxation by pair annihilation or drifting out of the system cannot account for the observed $\uf$ dependence of the saturated vortex number, thus supporting our early-time coarsening description of the defect saturation for fast quench. 

\begin{figure}
\includegraphics{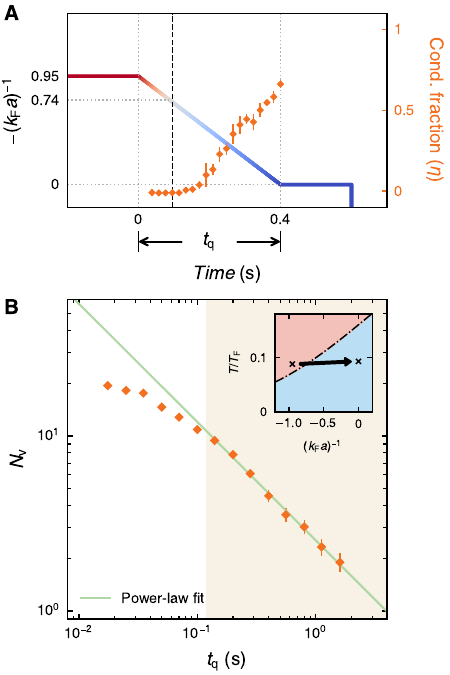}
\caption
{
\textbf{Interaction quench.}
(\textbf{A}) Illustration of interaction strength quench protocol. 
By tuning the Feshbach magnetic field, we quench the interaction parameter ($-(\kf a)^{-1}$) from 0.95 to 0 for a variable time $\tqq$ \cite{supp}.
The solid data points show the condensate fraction ($\eta$) along a quench trajectory with $\tqq=0.4$~s.
The black dashed line indicates the point of continuous phase transition ($-(\kf a)^{-1}=0.74$).
Each data point is an average of at least 5 experimental realizations.
The error bars denote the standard deviation of the measurements.
(\textbf{B}) Interaction strength quench time ($\tqq$) versus average number of vortices ($\nv$).
The inset depicts the quench trajectory along the phase diagram. 
The solid line represents a power-law fit in the shaded region.
Each data point is an average of 40 experimental realizations. 
The error bars denote the standard error of the mean.
}
\end{figure}

We now turn our attention to interaction quench. 
In a spatially uniform trap with the same disk geometry, we prepare a thermal cloud at a magnetic field of $B=940$~G, where $a<0$ and the system is on the Bardeen--Cooper--Schrieffer (BCS) side of the BCS--Bose--Einstein condensate (BEC) crossover (Fig. 1A). 
The initial number of atoms per spin state is $7\times10^{5}$ and the vertical trap frequency is kept at $\omega_{z}=2\pi\times$490 Hz throughout the quench. 
By linearly decreasing the Feshbach magnetic field to $B=830$~G over a variable time $\tqq$, the interaction parameter is quenched from $-(\kf a)^{-1} = 0.95$ to 0 (Fig. 3A) \cite{supp}. 
The diamonds in Fig. 3A show the condensate fraction ($\eta$) measured along a slow quench trajectory with $\tqq=0.4$~s. 
At the end of the quench, when the unitary regime is reached, the condensate fraction is around 70\% and the system is in the superfluid phase. After the quench, we hold the atoms for 0.2~s and take a ToF image of the atomic sample.

Figure 3B shows the average number of vortices after an interaction quench for different quench times $\tqq$.
In the slow quench regime of $0.1~\textrm{s}<\tqq<2~\textrm{s}$, the average number of vortices follows a scaling behavior with the quench time.
The power-law exponent is determined to be 0.68(2), which is in excellent agreement with the temperature quench results.
This is expected since the critical nature of the superfluid phase transition does not change regardless of which system parameter is adjusted for the transition, unless it involves a symmetry breaking field, such as spin population imbalance \cite{supp, Warner}. 
Thus, our observation that both the temperature and the interaction quench share the same $\akz$ is a demonstration of universal critical dynamics based on scale invariance. 
Additionally, the defect saturation behavior is observed to begin at $N_\textrm{v}\sim10$ for both quenches, implying a universal saturation level of the defect density in phase transition dynamics, as suggested in recent experiments \cite{Ko,Goo1,Rabga}.


\begin{figure}
\includegraphics{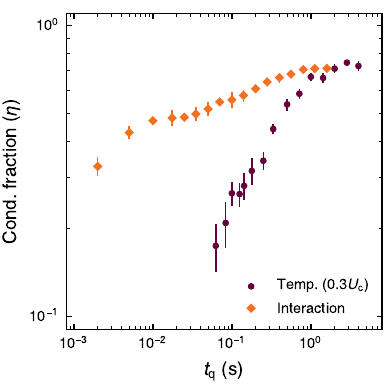}
\caption
{
\textbf{Condensate fraction versus quench time.}
The condensate fraction ($\eta$) as a function of quench time ($\tqq$) is shown for both temperature ($\uf=0.3\uc$) and interaction quench.
The condensate fraction was measured right after the end of the quench with the 0.2-s hold time omitted.
Each data point is an average of at least 5 experimental realizations.
The error bars denote the standard deviation of the measurements.
}
\end{figure}

It is evident that the microscopic nature of condensate formation is fundamentally different in the two cases, as demonstrated by the distinct timescales for condensate formation.
Figure 4 shows the condensate fraction for temperature and interaction quench right after the end of the quench with the 0.2-s hold time omitted.
For a temperature quench with $\tqq<0.3$~s, the condensate fraction already starts to decrease below 0.4.
In contrast, for an interaction quench, the condensate fraction at the end of the quench reaches below 0.4 only for the fastest 2-ms quench.
The fast timescale for condensate formation in an interaction quench agrees with a recent experiment of a weakly attractive Fermi gas quenched to the unitary regime \cite{Dyke}, where the timescale for pair formation was shown to be a few $\frac{\hbar}{E_\textrm{F}}$ with $\hbar$ being the reduced Planck constant and $E_\textrm{F}$ the Fermi energy, giving an order of 100 $\mu$s for our sample condition. 


\begin{figure}
\includegraphics{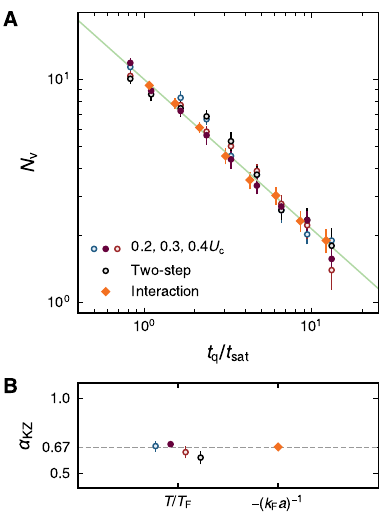}
\caption
{
\textbf{Universal KZ scaling.}
(\textbf{A}) Average number of vortices ($\nv$) as a function of normalized quench time ($\tqq/t_\textrm{sat}$) for the KZ scaling regime.
$t_\textrm{sat}$ represents the time at which $\nv=10$ from the power-law fits of temperature (Fig. 2B) and interaction (Fig. 3B) quench.
The solid line is a guide for the eye with an exponent of 0.67.
(\textbf{B}) KZ exponent ($\akz$) obtained for different quench protocols and parameters.
The error bars denote the standard deviation of the scaling exponent obtained from a power-law fit.
}
\end{figure}
We summarize our main results for the slow quench regime in Fig. 5.
The average number of vortices as a function of quench time for different quench protocols and parameters are encapsulated in Fig. 5A. 
By setting $t_\textrm{sat}$ to be the time at which $\nv=10$ in the power-law fits of temperature (Fig. 2B) and interaction (Fig. 3B) quench, we have scaled the horizontal axis to accommodate the different time scales involved in the quench dynamics.
Figure 5B shows the power-law exponent for different quench protocols and parameters. 


This work successfully realizes the gedanken experiment Zurek conceptualized four decades ago for liquid $^\textrm{4}$He.
Strongly interacting Fermi gases, owing to their low mass density, typically feature more than 8 orders of magnitude longer collision time than liquid $^\textrm{4}$He.
The large sample size and slow collision rate combined offer a time window for confirming the power-law relation between the average number of quantum vortices and the quench time.
An important extension of this work would be to probe the mean-field BCS transition for $-(\kf a)^{-1} \gg 1$, where $\akz$ is expected to be 0.5 \cite{Zurek, Kardar}.
Furthermore, the rich phase diagram in the presence of spin population imbalance suggests that near the tricritical point the critical behavior may exhibit a clear departure from the typical lambda transition, which will serve as a benchmark for critical phenomena of multicritical points \cite{Griffiths, H&H, Siggia3, Folk}.


\section{Acknowledgements}
We thank Eun-Gook Moon for helpful discussions and Jee Woo Park for helpful discussions and a critical reading of the manuscript.
This work was supported by the National Research Foundation of Korea (Grant No. NRF-2019H1A2A1074494, NRF-2023R1A2C3006565) and the Institute for Basic Science in Korea (Grant No. IBS-R009-D1).


\vspace*{-10pt}

\bibliographystyle{Science} 

\makeatletter
\renewcommand\@biblabel[1]{#1.}
\makeatother

\bibliography{main}

\onecolumngrid
\pagebreak


\begin{center}
\textbf{
Supplementary Materials for\\[4mm]
\large Observation of universal Kibble--Zurek scaling in an atomic Fermi superfluid}\\
\vspace{4mm}
{Kyuhwan Lee$^{1,2}$, 
Sol Kim$^{1,2}$, 
Taehoon Kim$^{1}$, 
Y. Shin$^{1,2,3,*}$}\\
\vspace{2mm}
{\em \small
$^1$\mbox{Department of Physics and Astronomy, Seoul National University, Seoul 08826, Korea}\\
$^2$\mbox{Center for Correlated Electron Systems, Institute for Basic Science, Seoul 08826, Korea}\\
$^3$\mbox{Institute of Applied Physics, Seoul National University, Seoul 08826, Korea}\\[2mm]}
{\small$^\ast$ Corresponding author. E-mail: yishin@snu.ac.kr}\\
\end{center}

\setcounter{equation}{0}
\renewcommand{\theequation}{S\arabic{equation}}

\setcounter{figure}{0}
\renewcommand{\thefigure}{S\arabic{figure}} 


\section{Sample preparation}

\noindent
Details of our experimental setup for preparing a strongly interacting Fermi gas of $^{6}$Li are described in Refs.~\cite{Park, Ko}.  
We start by loading $^\textrm{23}$Na and $^\textrm{6}$Li atoms in our dual-species magneto-optical trap.
After transferring both atomic species to an optically-plugged magnetic quadrupole trap, $^\textrm{23}$Na atoms are evaporatively cooled by using a radio-frequency (rf) knife during which $^\textrm{6}$Li is sympathetically cooled to quantum degeneracy.
Then the atoms are transferred into an oblate optical dipole trap (ODT) whose aspect ratio is given by 110:1.
By shining a resonant light, $^\textrm{23}$Na atoms are removed while $^\textrm{6}$Li atoms remain.
The $^\textrm{6}$Li atoms in $\ket{F=3/2, m_{F}=3/2}$ are first transferred to the $\ket{1}=\ket{F=1/2, m_{F}=1/2}$ state by applying a rf Landau--Zener (LZ) sweep centered at a DC magnetic field of 3~G, after which we increase the magnetic field to 870~G.
Later, a second LZ pulse is applied by changing the magnetic field from 870~G to 899~G so as to create a balanced mixture of $\ket{1}$ and $\ket{2}=\ket{F=1/2, m_{F}=-1/2}$.
Subsequently, dependent on the quench protocol, the depth of the ODT and the Feshbach magnetic field are set to a designated value.

To create a planar homogeneous trap, we use a spatial light modulator (SLM) that is placed on the imaging plane of a 4f configuration.  
Initially, a laser light with a wavelength of 532 nm is shined on the SLM to construct an optical wall potential.
Next, we take an $\it{in}$-$\it{situ}$ absorption image so that the difference between the target uniform density and the current column density acts as an error signal.
An affine transformation provides a suitable mapping between the CCD camera pixels and SLM pixels that lie on two separate planes.
Based on the error signal and the mapping between the camera and the SLM, the phase information of the individual pixels on the SLM is automatically tuned so that the error is minimized.
This process is continued until the column density image closely resembles the desired uniform density.
As a result, the average relative variations in the central disk region of the sample are maintained within 9\% throughout the quench.


\section{Temperature quench}

\begin{figure}[t]
\includegraphics{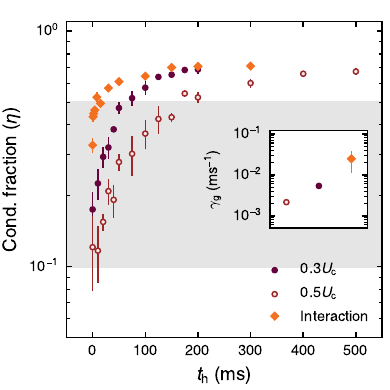}
\caption{
\textbf{Condensate fraction growth.}
Condensate fraction ($\eta$) as a function of the hold time ($t_\textrm{h}$) after quench.
The solid (open) circles show the growth of condensate fraction for a temperature quench with $\tqq=63$ ms ($\tqq=50$ ms) at $\uf=0.3\uc$ ($\uf=0.5\uc$). 
The solid and open circles share an identical quench rate for passing through the critical point.
The diamonds indicate the condensate fraction growth for an interaction quench with $\tqq=2$ ms.
The inset displays the initial growth rates ($\gamma_\textrm{g}$) calculated in the shaded region of the main figure.
Each data point is an average of at least 5 experimental realizations and its error bar denotes the standard deviation of the measurements.
}
\end{figure}

\noindent
Prior to initiating a temperature quench, the initial number of atoms per spin state is 1.0$\times$$10^{6}$ and the trap frequency along the tight confining direction is $\omega_{z}=2\pi\times$770 Hz.
A spatially uniform density profile is obtained across a disk of a diameter of 360 $\mu$m ($1/e$ diameter with respect to center density) at $U_{c}$.
The weak radial trapping potential provided by the ODT is slightly varied during the quench.
To compensate for the weakly changing radial trap potential, the power of the 532 nm laser light is ramped down by 29$\%$ in which the ramp speed is synchronized to the quench rate of the ODT and the start of the ramp down is when the ODT depth passes $U_{c}$.
After the end of the quench, the vertical trap frequency at 0.3$U_{c}$ is $\omega_{z}=2\pi\times$370 Hz.
For a final trap depth of 0.3$U_{c}$, the condensate fraction reaches its equilibrium value after a hold time of 0.2~s (Fig. S1).
Based on the condensate fraction growth data, we choose the hold time after the end of the quench to be 0.2~s, during which the 532 nm laser light is adiabatically turned off in order to enhance the visibility of quantum vortices in time-of-flight imaging.

Early-time coarsening characterizes a defect suppression effect during the early exponential growth of the IR modes after passing the critical point and when the notion of a quantum vortex is ill-defined.
The solid (open) circles in Fig. S1 show the growth of the condensate fraction for rapid quenching with $\tqq=63$ ms ($\tqq=50$ ms) when $\uf=0.3\uc$ ($\uf=0.5\uc$).
For $\uf=0.3\uc$, the initially low condensate fraction compared to the equilibrium condensate fraction suggests that the topological defects are not stably formed.
As a result, during the initial 50 ms, the rapid increase in the condensate fraction is expected to be accompanied by a large suppression in phase singularities.
At higher final temperature, as observed for $\uf=0.5\uc$, a longer condensate growth time would lead to stronger suppression \cite{Chesler}.
These observations support the trend of higher final temperature leading to lower saturated number of vortices in Fig. 2B.
Delayed condensate formation for fast temperature quench has similarly been observed in weakly interacting Bose--Einstein condensates (BECs) \cite{Goo1, Wolswijk}.
In contrast, the diamonds in Fig. S1 represent the condensate fraction as a function of hold time for an interaction quench with $\tqq=2$ ms.
The faster growth of the condensate suggests that the effect of early-time coarsening should be smaller for an interaction quench.  


\section{Interaction quench}

\begin{figure}
\includegraphics{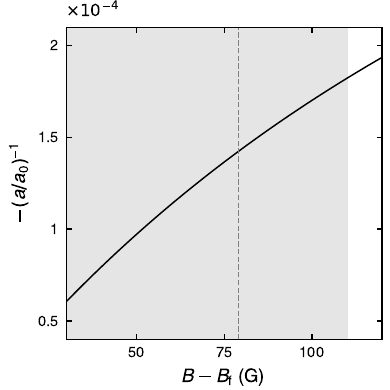}
\caption{
\textbf{Inverse $s$-wave scattering length versus Feshbach magnetic field.}
The solid line depicts the experimental data of the inverse $s$-wave scattering length ($a^\textrm{-1}$) as a function of the magnetic field ($B$) with respect to $B_\textrm{f}=830$ G \cite{Zurn}.
Here, $a_\textrm{0}$ denotes the Bohr radius.
The shaded region represents the quench range of the magnetic field in our experiment.
The dashed line indicates the point of continuous phase transition.
}
\end{figure}

\noindent
Initially, a normal gas of a balanced mixture of the two hyperfine states, $\ket{1}$ and $\ket{2}$, is prepared at a Feshbach magnetic field of 940 G.
The number of atoms per spin state is 7$\times$$10^{5}$ at a trap frequency of $\omega_{z}=2\pi\times$490 Hz.
The depth of the ODT corresponds to 0.5$U_{c}$, where $U_{c}$ is the critical depth at which the normal gas undergoes a superfluid transition in the temperature quench at the unitarity.
As in the temperature quench, we prepare a spatially homogeneous trap with the same diameter using an SLM.
For the quench of the interaction, we access the broad magnetic Feshbach resonance between the two hyperfine states, $\ket{1}$ and $\ket{2}$, which is centered at $B_\textrm{0}=832$ G.
The Feshbach magnetic field is tuned from $B_\textrm{i}=940$~G to $B_\textrm{f}=830$ G with varying quench times, during which the attractive Fermi gas transitions into a superfluid at $B_\textrm{c}=909$ G.
After the end of the quench, we apply a hold time of 0.2~s so that for even the fastest quench, the condensate fraction would reach its equilibrium value as shown in Fig. S1.
Identical to the thermal quench, the 532 nm laser light is adiabatically turned off to obtain maximum visibility of quantum vortices.

KZ mechanism deals with the rate at which the system passes through the critical window of freeze-out.
In order to translate the quench time of the magnetic field to the quench rate of the interaction parameter, a linear relation between the Feshbach magnetic field and the interaction parameter is required at the vicinity of the critical region.  
A broad magnetic Feshbach resonance between the two states, $\ket{1}$ and $\ket{2}$, can be modeled as $a=a_{bg}[1+\Delta (B-B_{0})^{-1}]$, where $a_{bg}$ is the background scattering length, $\Delta$ is the width of the resonance, and $B$ is the Feshbach magnetic field \cite{Zurn}. 
For the interaction quench, the resonance width (262 G) is much larger than our quench depth (110 G).
Consequently, for $B_\textrm{i}\geq B\geq B_\textrm{c}$, the Feshbach magnetic field and the inverse $s$-wave scattering length retain its linearity, as shown in Fig. S2.


\section{Determining $\akz$}

\noindent
The main quantitative model that the KZ mechanism establishes is the power-law behavior of the defect number as a function of the quench time.
A leveling off of the number of defects for fast quench times suggests that a power-law exponent should be extracted from only the slow quench regime.
To this end, we use a chi-square test ($\chi^{2}_{\nu}$) as a criterion for choosing the window for scaling as shown in Fig. S3 \cite{Rabga}.
Starting from slow quenches, the region of scaling is extended towards the fast quench regime until the $\chi^{2}_{\nu}$ sees a departure from its minimum value.
The scaling region for temperature and interaction quench had been chosen under this criterion, as highlighted by the shaded regions in Figs. S3A and S3B.
The point of increase in $\chi^{2}_{\nu}$ also coincides with the quench times at which $\akz$ begins to exhibit a trend of decreasing value.

\begin{figure}[h]
\includegraphics{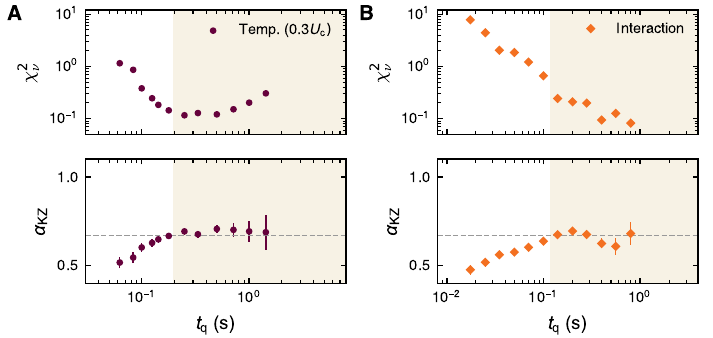}
\caption{
\textbf{Determination of the KZ scaling region.} 
The exponent $\akz$ was determined from a power-law fit to the data points for $\tqq\geq t_\text{L}$.
Measurement results of $\akz$ and the $\chi^{2}_{\nu}$ statistics as functions of the lower bound $t_\text{L}$ of the fitting region 
(\textbf{A}) for the temperature quench with $\uf=0.3\uc$ and (\textbf{B}) for the interaction quench. 
The error bars for $\akz$ denote the standard deviation of the scaling exponent obtained from the power-law fit.
The dashed line is a guide for the eye with $\akz=0.67$.
The shaded region indicates the KZ scaling region. 
}
\end{figure}


\section{Model $F$: Critical dynamics of superfluid transition}

\noindent
Model $F$ was first introduced to characterize the dynamic critical behavior of the lambda transition in liquid helium \cite{Siggia1, Siggia2}.
It successfully accounts for the divergence of second sound diffusion and thermal conductivity near the lambda transition.
In the following, we briefly sketch the model.

The singular part of the partition function $Z$ and the free energy $F_{0}$ are given as
\begin{equation}\label{eq: free energy}
\begin{aligned}
Z &\coloneqq \int D\psi Dm \exp\left[-F_{0}\left(\psi, m\right)\right] \\
F_{0}\left(\psi, m\right)&= \int d^3x\left(\frac{\lambda}{2}\vert\psi\vert^2+u\vert\psi\vert^4+\frac{K}{2}\left\vert\nabla\psi\right\vert^2+\frac{\chi^{-1}}{2}m^{2}+\gamma m\vert\psi\vert^2 \right).
\end{aligned}
\end{equation}
Eq. \ref{eq: free energy} consists of two slow variables, the order parameter $\psi$ and entropy density $m$, whose hydrodynamics is expressed by a generic Langevin equation.
Here, $\psi(x)$ is complex, $m(x)$ is real and the coupling constants are assumed to be real.
Notably, by integrating out $m$, we recover the static scaling behavior of the order parameter equivalent to that of the 3D XY model.
With the addition of mass density $\rho$ and mass current density $j$, the entire two-fluid hydrodynamics of a superfluid can be described.
However, near a continuous phase transition the dynamics of $\rho$ and $j$ are fast, as the temperature dependence of the speed of the first sound and its diffusion is weakly varying \cite{H&H2, Patel}.
Consequently, $\rho$ and $j$ can be ignored in the description of the critical dynamics of the superfluid transition. 

The Langevin equations that describe the motion of the slow variables are 
\begin{subequations}
\begin{align}
\frac{\partial \psi}{\partial t} &= -2\Gamma\frac{\delta F}{\delta \psi^{*}}-ig\psi\frac{\delta F}{\delta m}+\theta_{\psi}\label{eq: langevin psi}\\
\frac{\partial m}{\partial t} &= \Lambda\nabla^{2}\frac{\partial F}{\partial m}+2g~\textrm{Im} \left(\psi^{*}\frac{\delta F}{\delta \psi^{*}}\right)+\theta_{m},\label{eq: langevin m}
\end{align}
\end{subequations}
where the gaussian noise terms $\theta_{\psi}$ and $\theta_{m}$ satisfy
\begin{subequations}
\begin{align}
\langle \theta_{\psi}\left(x, t\right) \rangle &= \langle \theta_{m}\left(x, t\right) \rangle = 0 \label{eq: noise 1}\\
\langle \theta_{\psi}\left(x, t\right)\theta_{\psi}^{*}\left(x',t'\right)\rangle &= 4\Gamma\delta^{3}\left(x-x'\right)\delta\left(t-t'\right) \label{eq: noise 2}  \\
\langle \theta_{m}\left(x, t\right)\theta_{m}\left(x',t'\right)\rangle
&= -2\Lambda\nabla^{2}\delta^{3}\left(x-x'\right)\delta\left(t-t'\right).\label{eq: noise 3}
\end{align}
\end{subequations}
The coefficient $\Gamma$ is complex while $g$ and $\Lambda$ are both real. 
We are interested in the dynamics of the superfluid order parameter near a continuous phase transition. 
As a result, we assume $\psi$ to be not conserved while $m$ is conserved, both of which are indicated in the coefficients of the noise correlations in Eqs. \ref{eq: noise 2} and \ref{eq: noise 3}.
Another important feature of the Langevin equations are the second terms in the right hand side of Eqs. \ref{eq: langevin psi} and \ref{eq: langevin m}.
These terms are included to account for the role of the local chemical potential as a generator of time evolution of the complex order parameter.
This is in contrast to model $C$, which exhibits a different dynamic scaling behavior due to the absence of these terms.

The dynamic critical exponent $z$ dictates the scaling behavior of the relaxation time $\tau$ with respect to the wavenumber $k$ for long wavelengths ($k^{-1}\gtrsim\xi$) as $\tau\sim k^{-z}$.
A renormalization group (RG) calculation of model $F$ based on $\epsilon$-expansion shows that $z$ is entirely determined by static critical exponents as $z=\frac{3}{2}+\frac{\Tilde{\alpha}}{2\nu}$, where $\Tilde{\alpha}=\max(\alpha,0)$ \cite{Siggia1, Siggia2, H&H2}. 
Here, $\alpha$ is the critical exponent describing the behavior of the specific heat $C_\textrm{p}$ near the critical point as $C_\textrm{p}\sim\left|\lambda\right|^{-\alpha}$.
Near the lambda transition in liquid helium, measurements of specific heat in a zero-gravity environment have shown $\alpha=-0.0127(3)$ such that the specific heat is finite at the transition \cite{Lipa}.
As a result, in a superfluid phase transition, we expect $z=1.5$.
The calculations based on RG is consistent with the approach based on a dynamic scaling hypothesis which is derived from the onset of second sound for below $T_{c}$ \cite{H&H2}.
An important consequence of the dynamic scaling behavior is the divergence of second sound diffusivity and thermal conductivity near the lambda transition.

In a continuous phase transition process, a relevant variable of the order parameter is $\lambda$ (Eq. S1), which is given as a function of the microscopic properties of the system and macroscopically constrained by the thermodynamic variables that describe the state.
For our experiment, as we linearly tune a thermodynamic quench parameter, $\lambda$ is also expected to linearly ramp down from positive to negative so that the system undergoes a continuous phase transition to a superfluid state.
In the absence of a symmetry breaking field, for example a spin population imbalance, we expect the KZ exponent to be uniquely determined independent of the thermodynamic quench parameter.
In contrast, the microscopic timescale governing condensate formation may feature an orders of magnitude difference as shown in Fig. 4. 


\section{Variance-to-Mean ratio of the number distribution of quantum vortices}

\begin{figure}
\includegraphics{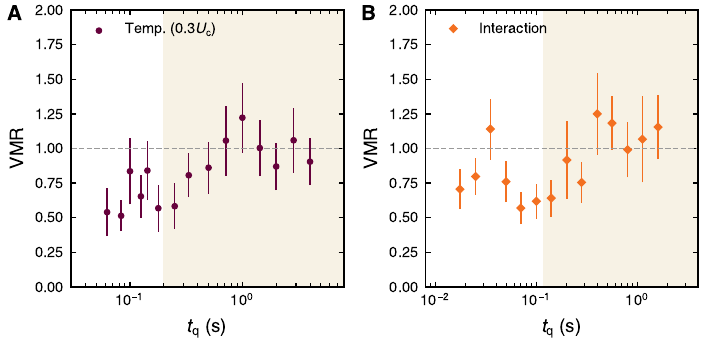}
\caption{
\textbf{Variance-to-mean ratio (VMR) of the vortex number}
(\textbf{A}) VMR of the vortex number for the single linear quench of temperature with $\uf=0.3 \uc$ is shown as a function of quench time.
(\textbf{B}) VMR of the vortex number for the interaction quench.
The error bars denote the standard error of the VMR.
}
\end{figure}

\noindent
The stochastic nature of defect formation can be summarized by a poisson distribution of the number of defects \cite{Gomez}.
A consequence of the poisson distribution is that the variance-to-mean ratio (VMR) converges to 1 in the limit of large number of trials.
This has recently been verified for a weakly interacting Bose gas confined in a harmonic trap by taking a large set of images per data point \cite{Goo1}.
Figure S4 shows the VMR of the defect number distribution for different quench parameters. 
In the scaling regime (shaded region), the VMR lies close to 1 within our error bars.
This suggests that the observed quantum vortices reflect the randomness inherent in independent domain formation, excluding destructive collisions of quantum vortices that lead to a suppression of VMR at the time of observation.  


\section{Relaxation of quantum vortices} 

\begin{figure}
\includegraphics{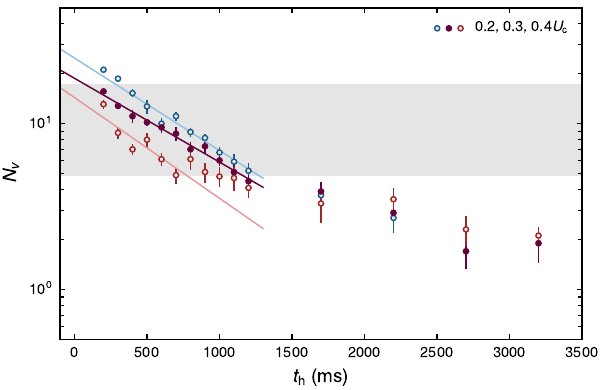}
\caption{
\textbf{Vortex relaxation.}
The decay of vortex number ($\nv$) as a function of hold time ($t_\textrm{h}$) after a temperature quench.
The blue open, solid, and red open circles correspond to $\uf=0.2\uc, 0.3\uc, 0.4\uc$, repsectively.
The solid lines are exponential fits to the data sets in the shaded region.
Each data point consists of at least 9 experimental realizations.
The error bars denote the standard error of the mean.
}
\end{figure}

\noindent
After going through an initial exponential growth of the condensate, the topological defect decay procedure begins.
The decay of quantum turbulence in superfluids mediated by quantum vortices transforming into compressible sound waves has recently been observed in strongly interacting Fermi gases \cite{Kwon}.
Tracking the average number of quantum vortices as a function of hold time ($t_\textrm{h}$) for the fastest temperature quench in Fig. 2, the typical timescale involved in vortex decay for fast quenches has been measured (Fig. S5).
Solid circles indicate the decay of vortices at $\uf=0.3\uc$.
The blue (red) open circles show the decay of vortices at $\uf=0.2\uc$ ($\uf=0.4\uc$).
By fitting $\nv$ as a function of $t_\textrm{h}$ with exponential decay, $\nv=N_\textrm{0}e^{-\gamma_\textrm{v}t}$, the decay rate ($\gamma_\textrm{v}$) in the shaded region of Fig. S5 is measured to be $\gamma_\textrm{v}=1.3(1)~\textrm{s}^{-1}$, $1.2(1)~\textrm{s}^{-1}$, $1.4(3)~\textrm{s}^{-1}$ for $\uf=0.2\uc, 0.3\uc, 0.4\uc$, respectively.
The decay rates are identical within our error bars, suggesting that the large temperature dependence of the saturated number of vortices does not originate from vortex relaxation.
Furthermore, the typical timescale involved in vortex relaxation, two-step quench results, and the VMR of the number distribution of quantum vortices all suggest that the effect of vortex decay would be minuscule for the slow quench regime.

\end{document}